\documentclass{aa}
\usepackage{graphicx}
\begin{document}

  \title{
Abundance anomalies in hot horizontal branch stars of the galactic globular
cluster NGC~1904\thanks{Based on data collected with the ESO {\it Very Large
Telescope + UVES}, ~at the Paranal Observatory, Chile}
 	 }

  \author{D. Fabbian\inst{1,2},
  A. Recio-Blanco\inst{1,3},
  R. G. Gratton\inst{4}, G. Piotto\inst{1} }

  \offprints{D. Fabbian, \email{damian@mso.anu.edu.au} }

  \institute{ Dipartimento di Astronomia, Universit\`a di Padova,
  		Vicolo dell'Osservatorio 2, 35122 Padova, Italy
	\and Current address: Research School of Astronomy \& Astrophysics, Australian
  		National University, Mount Stromlo Observatory, Cotter Road, Weston
  		ACT 2611 Australia
	\and Current address: Dpt. Cassiop\'ee, UMR 6202, Observatoire de la C\^ote d'Azur,
       BP 4229, 06304 Nice Cedex 4, France
	\and Osservatorio Astronomico di Padova, Vicolo dell'Osservatorio 5, 35122 Padova,
  		Italy}

  \date{Received / Accepted }

  \abstract{
We present  abundance measurements,  based on high-resolution  optical
spectroscopic  data obtained with   the Ultraviolet and Visual Echelle
Spectrograph mounted   on Kueyen (Very  Large Telescope  UT2), for ten
stars in the extended blue horizontal branch  of the Galactic globular
cluster NGC~1904 (M79). In agreement with  previous findings for other
clusters,  we obtain normal  abundances for stars  cooler than $T_{\rm
eff}\sim 11\,000$~K, and largely anomalous abundances for hotter stars:
large He depletions, and overabundances of Fe,  Ti, Cr, P and Mn. The
abundances  of  Mg,  Si and  Ca are  roughly  normal, in the hot stars
as well as in the cooler ones.
This  abundance  pattern  can be attributed   to  the onset  of
diffusion   and to radiation pressure  in  the stable atmospheres of hot
horizontal branch stars.

 }

  \authorrunning{D. Fabbian et al.}
  \titlerunning{Abundance Analysis of BHB stars in NGC~1904}
  \maketitle

\section{Introduction}

Sandage \& Wildey (1967) and van den Bergh (1967) first noticed that
the distribution of stars along the horizontal branch (HB) of globular
clusters (GCs) cannot be described as a simple uniparametric function
of metal abundance. In spite of intensive efforts since then, the
additional ({\it second}) parameter has not yet been clearly
identified; while numerous candidates have been proposed, none of them
is sufficient to justify the whole pattern of observations. The issue
is likely complicated by several factors. In particular, the
distribution of stars along the warmest side of the horizontal branch
(i.e.  stars bluer than the RR Lyrae instability strip, hence BHB) is
very complex, exhibiting a large variety of shapes, from short, stubby
horizontal branches (like e.g. in NGC~6397), to very long blue tails
(like e.g. in NGC~1904).

The distribution of stars along the HB of a GC is essentially
determined by their masses: the smaller the mass, the bluer its
colour. Since within a GC, stars have virtually the same age and
chemical composition and hence possibly the same mass at the turn-off
(see however D'Antona et al. 2002), the colour of stars along the HB
should be determined by the amount of mass lost by each star before
reaching the HB (the larger the mass lost, the bluer the colour on the
HB). While it seems quite reasonable to assume that mass loss occurs
more easily in extended and luminous red giants, the mechanisms
driving mass loss are far from being well understood, in particular in
the case of metal-poor, GC stars. Any possible observation relevant to
this topic might help to constrain these mechanisms.

Core rotation (Mengel \& Gross 1976; Sweigart \& Catelan 1998) is a
reasonable candidate for enhanced mass loss along the red giant branch
(RGB). In fact, core rotation reduces pressure in the helium cores,
delaying the helium flash that terminates evolution along the RGB.
Stars with rotating cores may then evolve to brighter luminosities,
and have the possibility to lose more mass.  We would then expect that
descendants of RGB stars with fast core rotation would inhabit the
bluest part of the HB. Some support for this hypothesis was given in
the '80s by observation of rotating BHB stars in M13 (a cluster with a
bluer than expected HB), but not in M3, by Peterson et al. (1983) and
Peterson (1983, 1985a, 1985b). In recent years, with the advent of
8-10~m class telescopes, this issue has been more extensively
revisited by Behr et al.  (1999, 2000a, 2000b; revised in Behr 2003a,
2003b) and more recently by our group (Recio-Blanco et al.
2002). These additional observations revealed a much more complicated
pattern than expected. Fast rotating stars are present also among
field HB stars (i.e., this is not an effect limited to globular
cluster stars, like the O-Na anticorrelation, see e.g. Kraft 1994),
and are present in most clusters (even M3, according to Behr 2003a).
They are however found only among the stars cooler than $\sim
11\,500$~K. This temperature value plays an important role in the HB
picture: stars warmer than this (and cooler than about 20\,000~K) have
anomalously bright $u$\ magnitudes in Str\"omgren photometry (Grundahl
et al. 1998, 1999); also, gravities derived for these stars from the
wings of Balmer and helium lines are systematically lower than
predicted by evolutionary models (Moehler et al.  1995, 2003). As
shown by several authors (Grundahl et al.  1999; Moehler et al.  1999;
Hui-Bon-Hoa et al.  2000), this can be attributed to the onset of
strong effects of levitation due to microscopic diffusion and
radiation pressure that can be seen in stars lacking significant
sub-atmospheric convection zones (Sweigart 2002).  When microscopic
diffusion and radiation pressure are not overwhelmed by convection,
the atmospheres of warm stars are expected to be strongly depleted in
He (Greenstein et al. 1967; Michaud et al.  1983), and to exhibit
large overabundances of elements like Fe, Ti and P (Turcotte et
al. 1998; Richer et al. 2000).  All these features are indeed observed
(Glaspey et al. 1989; Moehler et al. 1999; Behr 2003a).

It should be noted that most of these results - in particular those
related to the chemical composition - have been obtained only for a
few GCs, mainly by a single group.  It would be useful to extend such
observations to more clusters, performing an independent analysis.  In
this paper we present a new analysis of the chemical composition of 10
BHB stars in the southern cluster NGC~1904 (M79).  NGC~1904 is a
cluster of intermediate metal abundance ([Fe/H]=-1.59 according to
Kraft \& Ivans 2003, using Kurucz 1994 models), with an extended very
blue HB: these properties are similar to those of M13, one of the
clusters studied by Behr (2003b). The observational material obtained
by us for NGC~1904 was originally intended only for measurements of
the rotational velocities: hence the spectra have a low
signal-to-noise ratio ($S/N$), although high resolution.  However, we
will show that abundances good enough for the purposes of the present
discussion could be obtained even from this material.

\label{intro}

\section{Observations and reduction}
\label{observation}

\begin{figure*}[t]
\centering
\includegraphics[width=18cm,height=7cm]{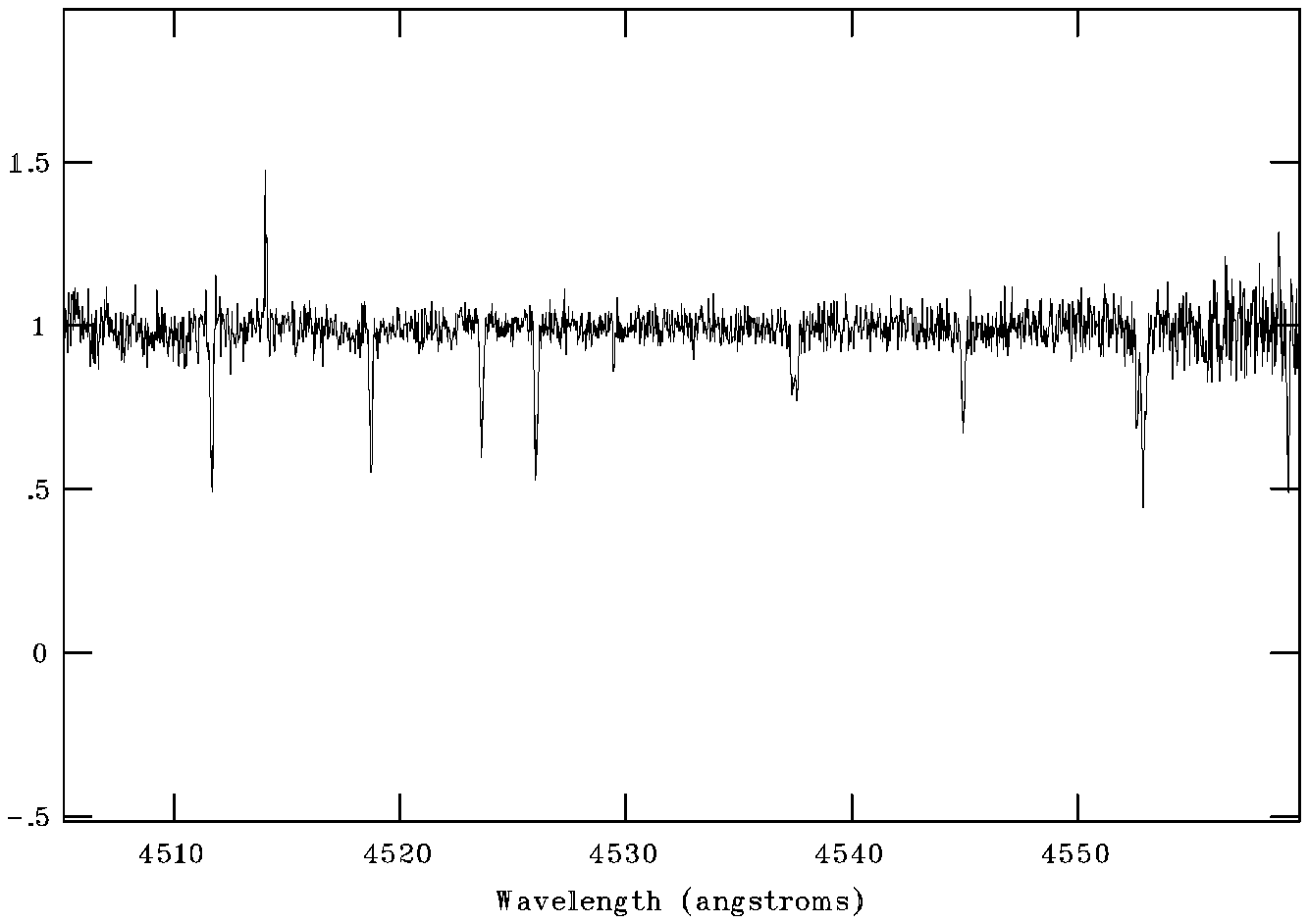}\\
\vspace{0.8cm}
\includegraphics[width=18cm,height=7cm]{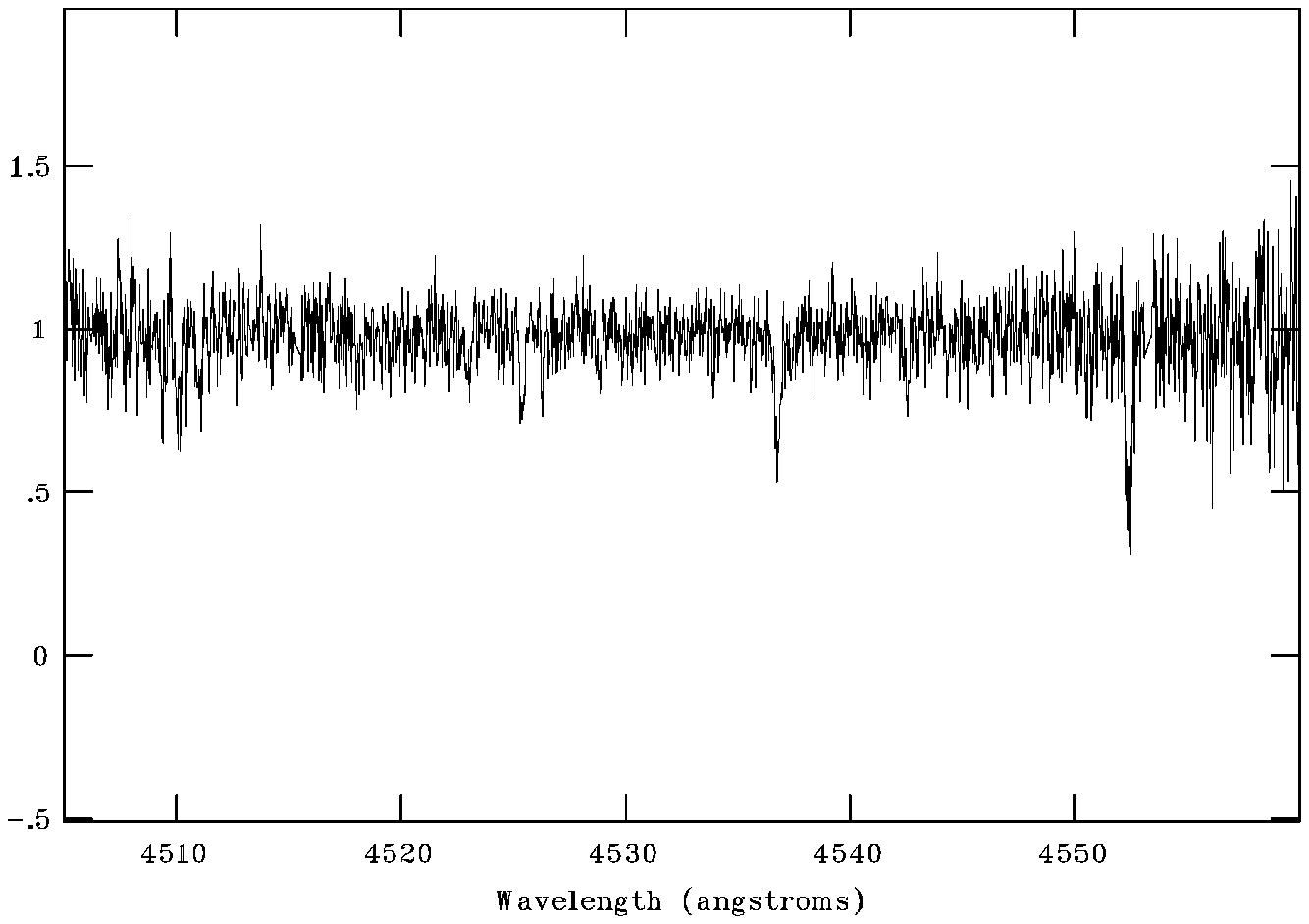}
\caption{Spectral sections of the stars 392 (top panel), and 209
(bottom panel)}
\label{f:fig1}
\end{figure*}

A total of 21 stars were observed in NGC~1904; however, spectra for
only 10 of these stars were deemed of good enough quality for
abundance analysis, the criteria being the $S/N$\ of the spectra and
the rotational velocity of the stars (only slow rotators were
considered).  We selected our targets from the F439W and F555W
HST-WFPC2 photometry by Piotto et al. (2002).  In addition, we have
also identified the NGC~1904 targets in the Str\"omgren {\it u, y}
photometry by Grundahl et al. (1999) and the Johnson {\it U, V}
photometry by Momany et al. (2004). In general, targets lie in the
low-density outskirts of the GC, to avoid contamination from other
stars.  The spectra were obtained during 2 observing runs: July
30--August 2 2000 and January 19--23 2001.

The observations were carried out with the UVES echelle spectrograph,
mounted on the Unit 2 (Kueyen) of the VLT, and the 2K x 4K, 15 $\mu$m
pixel size blue CCD, with a readout noise of 3.90 e$^{-}$ and a
conversion factor of 2.04 e$^{-}$/ADU. The UVES blue arm, with a
spectral coverage in the 3730-5000 $\stackrel {\circ} {\rm A}$ range,
was used.  Combined with a slit width of 1.0 arcsec, we achieved a
resolution of R$\sim$ 40\,000 ($\delta \lambda \sim 0.1
\stackrel {\circ} {\rm A}$, $\delta v \sim$ 7.5 km/s, see the UVES user
manual by Kaufer et al. 2003).

The exposure times ranged from 800 s (for V $\sim$ 16) to 4000 s (for
V $\sim$ 17.5). We generally limited individual exposure times to 1500
or 2000 seconds, to minimize cosmic ray accumulation, and then coadded
individual frames obtained for each star. The typical signal to noise
ratios of the coadded spectra are about 20 to 30 per resolution
element, but in a few cases they are as good as $S/N\sim$40-60. See
Fig.~\ref{f:fig1} for a visual example of the quality of the target
spectra.

The spectra were extracted using standard IRAF procedures. For the
wavelength calibration we fitted 3$^{rd}$ order polynomials to the
dispersion relations of the ThAr calibration spectra, which resulted
in residuals of $\leq$ 3$\cdot$10$^{- 4}$ $\stackrel {\circ} {\rm
A}$. Finally, each spectrum was normalized using a 5th-degree
polynomial fitting.

\section{Abundance measurements}

We have performed an abundance analysis of 10 stars in M79. The
abundance measurements were performed using the program WIDTH3,
developed by R. G.  Gratton and adapted by D. Fabbian to temperatures
up to $T_{\rm eff} \simeq$ 20\,000~K (D.  Fabbian graduate
thesis). The procedure establishes the abundances of chemical species
by reproducing the observed equivalent widths.  Once a starting set of
values for the effective temperature, surface gravity and model metal
abundance is derived, an appropriate model for each star is obtained
from the grid of model atmospheres by Kurucz (1994) by interpolating
linearly in temperature and logarithmically to get the Rosseland
opacity, electronic and gaseous pressure, and density.  Continuum
opacity was obtained taking into account all important continuum
opacity sources for stars as hot as $T_{\rm
eff}=20\,000$~K. Collisional damping constants were computed with the
Uns$\ddot{\rm o}$ld formula.  The equation of transfer was then
integrated through the atmosphere at different wavelengths along the
line profile and theoretical equivalent widths of the lines computed
and compared with the observed ones.

\subsection{Oscillator strengths, line selection and solar abundances}

Only lines free of blends were considered in the analysis. The line
list was extracted from the solar spectrum tables by Moore et al.\
(1966), and the atomic line lists by Hambly et al. (1997) and Kurucz
\& Bell (1995), more appropriate for warmer stars.

For the determination of the oscillator strengths ({\it gf}s),
laboratory values were considered whenever possible.  For \ion{Fe}{i}
lines they were taken from papers of the Oxford group (Simmons \&
Blackwell 1982) and Bard \& Kock (1994). The {\it gf}s for the
\ion{Fe}{ii} lines were taken from Heise \& Kock (1990) and Biemont et
al.\ (1991), while {\it gf}s for the \ion{Mg}{i} lines came from
Gratton (2003). For the rest of the lines, the {\it gf}s were those
provided by Kurucz (1994) and the NIST atomic spectra database.

The solar abundances were taken from Grevesse \& Sauval (1998).

\subsection{Equivalent Widths}

Equivalent Widths ($EW$s) were measured on the unidimensional,
extracted spectral orders using an automatic routine within the ISA
package, prepared by R.  Gratton (see Bragaglia et al. 2001).  The
routine works as follows. First, a local continuum level is determined
for each line by an iterative clipping average of the 200 spectral
points centered on the line to be measured.  Second, the equivalent
width for each line from an extensive list is tentatively measured
using a Gaussian fitting routine. Measures are rejected according to
several criteria (e.g., if the central wavelength does not agree with
that expected from a preliminary measure of the geocentric radial
velocity or if the lines are too broad or too narrow). Third, the
measured lines are used to draw a relationship between equivalent
width and FWHM. Fourth, this relation is used to measure equivalent
widths again, using a second Gaussian fitting routine that has only
one free parameter (the equivalent width), since the central
wavelength is fixed by the measured geocentric radial velocity and the
FWHM is fixed by the relationship between equivalent width and
FWHM. Again measures are rejected if residuals are too large,
asymmetric, etc.

But for a few exceptions (\ion{Mg}{ii} and \ion{Ca}{ii} lines having
$EW>$100~m\AA\ ), only lines with 10$<EW<$100~m\AA\ were finally
considered. Lines with $EW<$30~m\AA\ were only used for the star 392,
the one with the best spectrum (see section 4.2 and
Figure~\ref{f:fig1}). The number of measured lines depends on the
$S/N$\ and on the star's metallicity. Generally, around 15 \ion{Fe}{i}
lines and 20
\ion{Fe}{ii} were  measured. Measured $EW$s for individual stars are
listed in  Table~\ref{t:tab1}, only available   in electronic form.

\begin{table}
\caption{The measured equivalent widths and corresponding uncertainties 
used in our abundance analysis. 
The table is available only in electronic form at the CDS. Column 1
lists the chemical elements of interest with their ionization stage;
Column 2 gives the corresponding wavelengths (in \AA); excitation
potentials (in eV) and $\log$ {\it gf}s are given in Column 3 and 4
respectively; the measured equivalent widths and corresponding
uncertainties (in m\AA) for the 10 stars are listed in the next
columns, with a few values being upper limits}
\label{t:tab1}
\end{table}

\begin{table}
\caption{Abundance errors for the star 392, due to uncertainties $\sigma(\Delta EW)$ in equivalent
width, $\sigma(\Delta T)$ in effective temperature, $\sigma(\Delta g)$ in $\log g$, $\sigma(\Delta \xi)$ in $\xi$ and $\sigma(\Delta z)$in [A/H]}
\begin{tabular}{llllll}
\hline
\hline
Element & $\sigma(\Delta EW)$& $\sigma(\Delta T)$ & $\sigma(\Delta g )$ &
$\sigma(\Delta \xi)$ & $\sigma(\Delta z)$ \\
       &$\pm 10.4$&$\pm 200$&$\pm 0.1$&$\pm 1$&$\pm 0.1$\\
       &(m\AA)    &(K)      &dex      &km/s   &   dex   \\
\hline
\ion{He}{i}   &  0.3  & 0.11 & 0.06 & 0.02 & 0.09 \\
\ion{Mg}{ii}  &  0.2  & 0.02 & 0.01 & 0.08 & 0.00 \\
\ion{Si}{ii}  &  0.2  & 0.02 & 0.03 & 0.03 & 0.01 \\
\ion{P}{ii}   &  0.2  & 0.02 & 0.03 & 0.05 & 0.01 \\
\ion{Ca}{ii}  &  0.4  & 0.05 & 0.04 & 0.22 & 0.01 \\
\ion{Ti}{ii}  &  0.2  & 0.09 & 0.00 & 0.05 & 0.02 \\
\ion{Cr}{ii}  &  0.2  & 0.04 & 0.02 & 0.03 & 0.02 \\
\ion{Mn}{ii}  &  0.2  & 0.03 & 0.02 & 0.02 & 0.01 \\
\ion{Fe}{i}   &  0.2  & 0.10 & 0.03 & 0.05 & 0.04 \\
\ion{Fe}{ii}  &  0.2  & 0.02 & 0.03 & 0.09 & 0.01 \\
\hline
\end{tabular}
\label{t:tab2}
\end{table}

Due to the quality of the spectra, the $EW$\ uncertanties were the
most important source of errors in the abundance determinations. This
error was evaluated by comparing the equivalent width measurements of
the stars 209 and 281.  These stars have almost equal atmospheric
parameters, with differences just slightly larger than the internal
errors. In addition, they do not present signs of diffusive effects in
their atmospheres. An analysis of 22 lines measured in both spectra
indicates that the mean difference between the two equivalent width
measurements is $-1.1$~m\AA\ (in the sense 281-209), with a standard
deviation of 14.7 m\AA.  Therefore, assuming that equivalent width
errors for both stars are the same, the error for each measurement is
14.7/$\sqrt{2}$=10.4~m\AA. If we compare this result with the value
calculated through the Cayrel (1988) formula: $$\Delta(EW) = 1.6
\sqrt{(w dx)}/(S/N)$$ with $w$\  being the full  width at half maximun
(FWHM) typical of the lines, in this case $\sim$0.12~\AA, and $dx$\
being the pixel size, about 0.03~\AA, we get $\Delta(EW)\simeq
10$~m\AA\ (assuming a $S/N\simeq$10 per pixel), which is more or less
what we obtained comparing the measurements for the stars 209 and
281. This result implies that identification of the correct continuum
level (that is neglected in the Cayrel formula) is not an issue here,
as expected since the spectra of BHB stars have very few absorption
lines.

Consequently, we calculated the abundance errors, derived from an
equivalent width uncertanty of 10.4~m\AA.  The resulting errors for
the star 392, $\sigma(\Delta EW)$, are shown in Column 2 of
Table~\ref{t:tab2}.

\subsection{Atmospheric parameters}

\begin{figure}
\centering
\includegraphics[width=7cm]{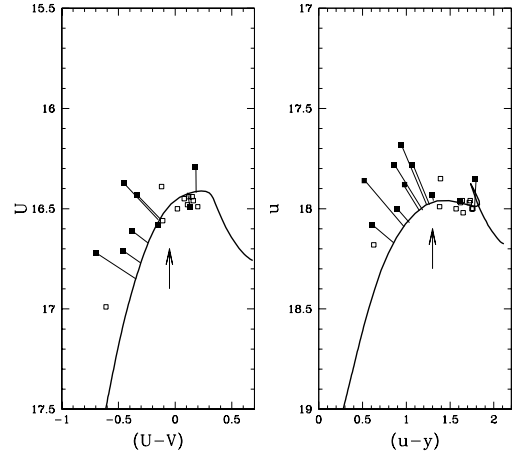}
\caption[] {Colour magnitude diagram for the target stars. Filled symbols are
stars for which abundance analysis was done; open symbols are those stars for
which only rotational velocities were obtained. Left panel: $U-(U-V)$\ colour
magnitude diagram from Johnson photometry by Momany et al. (2004). Right
panel: $u$,$(u-y)$\ diagram from Str\"omgren photometry by Grundahl et al.\
(1999). ZAHB models by Cassisi et  al. (1999) are plotted as continuous lines.
The location of the luminosity jump is marked with an arrow in both CMDs}
\label{f:fig2}
\end{figure}

\begin{table*}
\caption[]{Photometry for all the target stars in NGC~1904. The temperature values
derived from the $u$,$(u-y)$ Grundahl et al. (1999) CMD, $T_{\rm
effg}$, the corresponding corrected effective temperature values
$T_{\rm effgc}$ and the temperature derived from the Momany et
al. (2004) CMD, $T_{\rm effm}$}
\begin{tabular}{lllllllllll}
\hline
\hline
\textbf{ID} &  \textbf{y}  &  \textbf{b} & \textbf{v}  &  \textbf{u}  &
\textbf{V}  &  \textbf{B}  & \textbf{U} & \textbf{$T_{\rm effg}$} &
\textbf{$T_{\rm effgc}$} & \textbf{$T_{\rm effm}$} \\
 & & & & & & & & (K) & (K) & (K)\\
\hline
 243 & 16.23 & 16.29 & 16.47 & 17.96 & 16.29 & 16.31 & 16.44  & ~7700 & ~8684 & ~8500 \\
 209 & 16.06 & 16.14 & 16.34 & 17.85 & 16.11 & 16.16 & 16.29  & ~7500 & ~8537 & ~8400 \\
 294 & 16.24 & 16.31 & 16.49 & 18.00 & 16.30 & 16.32 & 16.46  & ~7600 & ~8611 & ~8600 \\
 295 & 16.37 & 16.41 & 16.56 & 18.02 &     &      &     & ~8000  &             &       \\
 289 & 16.32 & 16.37 & 16.52 & 17.96 & 16.37 & 16.36 & 16.45  & ~8100 & ~8976 & ~9000 \\
 275 & 16.25 & 16.31 & 16.48 & 17.97 & 16.32 & 16.32 & 16.44  & ~7700 & ~8684 & ~8800 \\
 297 & 16.46 & 16.47 & 16.57 & 17.85 & 16.51 & 16.43 & 16.39  & ~9300 & ~9853 & 10200 \\
 327 & 16.43 & 16.46 & 16.61 & 18.00 & 16.48 & 16.45 & 16.50  & ~8400 & ~9195 & ~9400 \\
 292 & 16.25 & 16.32 & 16.50 & 18.00 & 16.29 & 16.37 & 16.49  &       &       & ~8400    \\
 281 & 16.35 & 16.38 & 16.53 & 17.96 & 16.36 & 16.41 & 16.49  & ~8200 & ~9049 & ~8800 \\
 298 & 16.35 & 16.38 & 16.53 & 17.97 & 16.37 & 16.40 & 16.48  & ~8200 & ~9049 & ~8900 \\
 489 & 17.34 & 17.31 & 17.35 & 17.86 & 17.42 & 17.24 & 16.72  & 14500 & 13654 & 13000 \\
 354 & 16.64 & 16.67 & 16.79 & 17.93 & 16.73 & 16.67 & 16.58  & ~9800 & 10219 & 10600 \\
 469 & 17.11 & 17.10 & 17.16 & 18.00 & 17.17 & 17.03 & 16.71  & 12300 & 12046 & 12200 \\
 434 & 16.92 & 16.92 & 16.99 & 17.78 & 16.99 & 16.86 & 16.61  & 12300 & 12046 & 11600 \\
 389 & 16.71 & 16.70 & 16.77 & 17.78 & 16.77 & 16.65 & 16.43  & 11000 & 11096 & 11200 \\
 363 & 16.90 & 16.92 & 17.01 & 17.88 & 17.45 & 17.32 & 17.07  & 11800 & 11681 &       \\
 366 & 16.61 & 16.62 & 16.74 & 17.99 & 16.67 & 16.61 & 16.56  & ~9400 & ~9926 & 10200 \\
 555 & 17.55 & 17.51 & 17.56 & 18.18 & 17.60 & 17.44 & 16.99  & 13700 & 13070 & 13800 \\
 392 & 16.74 & 16.74 & 16.81 & 17.68 & 16.82 & 16.67 & 16.37  & 12100 & 11900 & 11500 \\
 535 & 17.47 & 17.47 & 17.54 & 18.08 & 18.07 & 18.14 & 17.50  & 13700 & 13070 &       \\
\hline
\end{tabular}
\label{t:tab3}
\end{table*}

Model atmospheres appropriate for each star were extracted from the
grid of Kurucz (1994). Figure~\ref{f:fig2} shows the projections of
the analyzed HB stars into the corresponding zero age horizontal
branches (ZAHBs) for the two adopted photometries. The position of the
Grundahl et al. (1999) luminosity jump is marked with a vertical
arrow. This feature clearly affects the fit of the models to the
observed HBs, especially in the $u$,$(u-y)$ CMD. $T_{\rm eff}$\ values
were determined from the position of the stars in the $u$,$(u-y)$ CMD
by Grundahl et al.\ (1999), and the position in a Johnson $U$,($U-V$)
CMD by Momany et al. (2004), using the ZAHB models by Cassisi et
al. (1999), see Fig.~\ref{f:fig2}. Those models were transformed from
the theoretical plane to the observational one using the
colour-effective temperature relation provided by Castelli et al.\
(1997a,b). Hereafter, we adopt a reddening value of $E(B-V)=0.01$,
from the Harris (1996) compilation. The value that could be obtained
using the maps of Schlegel et al. (1998) based on COBE-DIRBE data is
$E(B-V)=0.031$. If we had rather adopted this last value of the
interstellar reddening, the temperatures would have been $\sim 200$~K
warmer; this is less than the internal errors.

Table~\ref{t:tab3} presents the photometric parameters for all 21
stars in NGC~1904 for which we obtained spectra, with the
corresponding temperature values derived from the Grundahl et al.
(1999) $u$,$(u-y)$ CMD and the Momany et al. (2004) CMD, respectively
labelled $T_{\rm effg}$ and $T_{\rm effm}$. Only 10 of these stars,
selected among the slow rotating ones, were used in the abundance
analysis.

\begin{table}
\caption{Atmospheric parameters for stars in NGC~1904}
\begin{tabular}{lllll}
\hline
\hline
star  & $T_{\rm eff}$  & $\log g$ &$[$A/H$]$&$\xi$   \\
      &     (K)        &          &         &(km/s)  \\
\hline
209 & ~8469$\pm$200 & 3.2$\pm$0.2 & -1.4 & 2.4$\pm$1.0 \\
281 & ~8925$\pm$200 & 3.3$\pm$0.2 & -1.4 & 2.3$\pm$1.0 \\
354 & 10409$\pm$200 & 3.7$\pm$0.1 & -1.4 & 2.0$\pm$1.0 \\
389 & 11148$\pm$200 & 3.8$\pm$0.1 &  0.0 & 0.0$\pm$1.0 \\
363 & 11681$\pm$300 & 3.9$\pm$0.1 &  0.0 & 0.0$\pm$1.0 \\
392 & 11700$\pm$200 & 3.9$\pm$0.1 &  0.0 & 0.0$\pm$1.0 \\
434 & 11823$\pm$200 & 3.9$\pm$0.1 &  0.0 & 0.0$\pm$1.0 \\
469 & 12123$\pm$200 & 4.0$\pm$0.1 &  0.0 & 0.0$\pm$1.0 \\
535 & 13070$\pm$300 & 4.1$\pm$0.1 &  0.0 & 0.0$\pm$1.0 \\
489 & 13327$\pm$200 & 4.2$\pm$0.1 &  0.0 & 0.0$\pm$1.0 \\
\hline
\hline
\end{tabular}
\label{t:tab4}
\end{table}

To derive the final effective temperature of the programme stars, we
calculate the mean of Momany's temperatures and of Grundahl's
temperatures after transposing these last to Momany's scale. The final
error is the scatter of the temperatures involved in the mean. We
decided to convert the $T_{\rm effg}$ values to Momany's scale mainly
because the $T_{\rm effg}$ scale presented a number of problems. For
instance, the $T_{\rm effg}$ temperature for the coolest stars would
place them inside the RR~Lyrae instability strip, and the
corresponding metallicity for such a cool temperature would be lower
than what is reported for the RGB stars in the cluster.  On the other
hand, apart from this shift to cooler temperatures, probably due to
the ZAHB model fit, all of our targets were well identified in the
Grundahl et al. photometry, while 3 of the stars in Momany's
photometry were blended. For this reason, and to be sure of an
internal consistency in all the temperature measurements, we decided
to transform the $T_{\rm effg}$ values into Momany's scale, and then
calculate the mean, for each star, of the new $T_{\rm effg}$ obtained
and the $T_{\rm effm}$ value.

The relation between $T_{\rm effg}$\ and $T_{\rm effm}$\ was found by
averaging the best fit relations obtained assuming the two quantities
as independent variables, because internal errors in $T_{\rm effm}$\
and $T_{\rm effg}$\ were expected to be comparable. The final relation
is:
\begin{equation}
T_{\rm effgc} = (0.731 \pm 0.039) T_{\rm effg} + (3055 \pm 357)~~~{\rm K}
\end{equation}

where $T_{\rm effgc}$ is the corrected effective temperature derived
from the Str\"omgren {\it u, y} photometry by Grundahl et
al. (1999). Internal errors in these temperatures may be obtained by
comparing the differences between the temperatures obtained from the
two photometries. The r.m.s. average of these differences (taking into
account the two degrees of freedom lost because we corrected the
temperatures obtained from Grundahl's photometry to those obtained
from Momany's photometry) is 355~K; assuming that individual errors
are similar in the two determinations, the expected internal errors
are $\pm 177$~K for those stars with both determinations, and $\pm
251$~K for stars having only Grundahl's photometry. Hereafter, we will
round these values to $\pm 200$\ and $\pm 300$~K
respectively\footnote{Systematic errors have of course larger values
than these internal errors}, see Table~\ref{t:tab4}.

\begin{figure}
\centering
\includegraphics[width= 7cm]{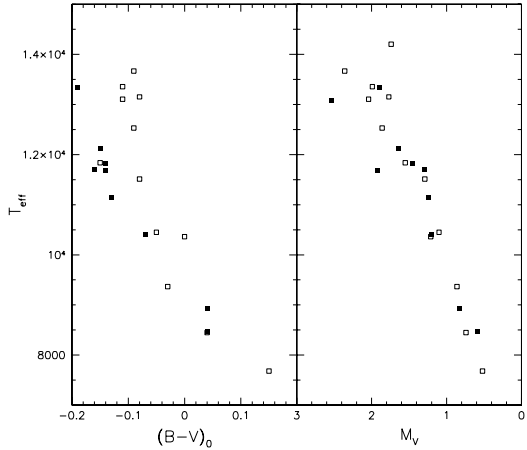}
\caption[]{Runs of $T_{\rm eff}$\ against the dereddened Johnson
$B-V$\ color (left panel) and absolute $M_V$\ magnitude (right
panel) for our program stars in NGC~1904 (filled symbols), and for
the stars in M13 analyzed by Behr (2003a) (open symbols)}
\label{f:fig3}
\end{figure}

In order to compare the effective temperatures adopted throughout this
paper with those adopted in other abundance analyses of BHB stars, we
plotted (see Fig.~\ref{f:fig3}, left panel) the runs of $T_{\rm eff}$\
against the dereddened Johnson $B-V$\ color and of the absolute $M_V$\
magnitude for both our program stars in NGC~1904 and the stars in M13
analyzed by Behr (2003a). The reddenings adopted here are those from
the Harris (1996) compilation.  There is a reasonable agreement
between the two sets for the coolest stars, while our temperatures for
the hottest ones are lower than those obtained by Behr for stars of
similar $B-V$\ colours. This difference is likely due to some
differences in the colour calibration; in fact, the agreement between
the two sets is excellent if absolute $M_V$\ magnitudes are considered
(see Fig.~\ref{f:fig3}, right panel).

Gravities are not well costrained by our spectra: in fact equilibrium
of ionization is subject to possible departures from LTE, and the
Balmer lines are too broad for reliable determinations of their
profile from our Echelle spectra. We have then derived a mean relation
between $\log T_{\rm eff}$\ and gravity ($\log g$) from Behr et al.\
(1999) measurements of blue HB stars in M13 that were expected to be
very similar to those in NGC~1904 on the basis of the colour-magnitude
diagram:
\begin{equation}
\log g = 4.83 \cdot \log T_{\rm eff} - 15.74
\end{equation}

\begin{figure}
\centering
\includegraphics[width= 7cm]{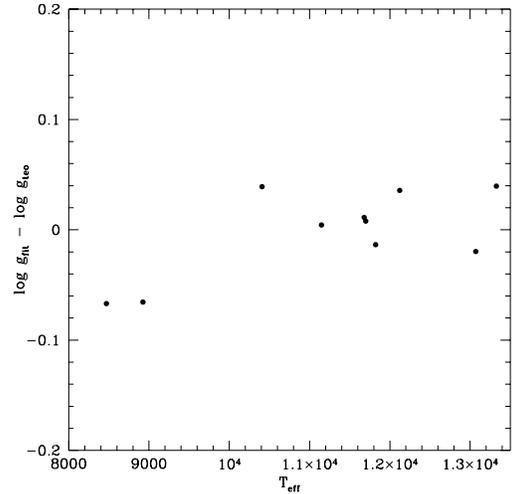}
\caption[]{Comparison between the gravities adopted in this paper
($\log g_{\rm fit}$) and those obtained from the location of the
stars in the colour-magnitude diagram, using theoretical models by
Cassisi et al. (1999: $\log g_{\rm teo}$)}
\label{f:fig4}
\end{figure}

Figure~\ref{f:fig4} compares the gravities obtained using this method
(and used by us in the abundance determinations), with those that
could be derived from photometric angular diameter. In the latter
case, a distance modulus of (m-M)$_V=15.59$ (Harris, 1996, in the
revised version of 2003) has been adopted for the cluster, and a
reasonable mass for the stars was estimated from HB models by Cassisi
et al. (1999) for the Johnson photometry, again using the location of
the stars in the colour-magnitude diagram.  The agreement between
these two values is very good, supporting the gravity estimates
adopted throughout this paper. Errors in the gravities due to errors
in the temperatures are of about 0.05~dex.

Microturbulent velocities $\xi$ might be derived by eliminating trends
of the derived abundances with expected line strength for some given
species. However, given the quality of our spectra, generally too few
lines could be measured, and with too large scatter, to significantly
constrain microturbulent velocity.  For the cooler stars, we relied on
Behr's analysis of M13 stars. From his analysis, we derived the
following relation between $\log T_{\rm eff}$\ and the microturbulence
velocity:
\begin{equation}
\xi = -4.7 \cdot \log T_{\rm eff} + 20.9 ~~~{\rm km/s}
\end{equation}
Uncertainties in these values for the microturbulent velocities can be
obtained from the scatter of individual values around this mean
relation: this is $\sim 1$~km/s. For the warmer stars, we adopted no
microturbulent velocity, as given by the analysis of star 392, which
has the best spectrum (see Section 4.2).

Metallicities were obtained varying the metal abundance [A/H] of the
model until it was close to the derived [Fe/H] value. The adopted
atmospheric parameters are listed in Table~\ref{t:tab4}. The errors.

Once $T_{\rm eff}$, $\log g$, $\xi$\ and atmospheric [Fe/H] were
determined, we calculated the mean abundance and dispersion for each
element. Since the abundances for each element depend upon the adopted
parameters, we recalculated the abundances varying each of these
parameters in turn, to test how much the abundances change.  The
resulting errors for the star 392 are shown in Table~\ref{t:tab2},
Columns 3 to 6.

\section{Results}
\label{results}

\begin{table*}
\caption{Abundances for stars in NGC~1904. The superscripts next to the abundance determinations indicate the number of lines used in each case}
\begin{tabular}{llllllllll}
\hline
\hline
star& log He           & log Mg          & log Si          & log P           & log Ca          & log Ti           & log Cr          & log Mn            & log Fe           \\
\hline
209 &     ...          &6.4$\pm$0.3$^{2}$&6.2$\pm$0.2$^{1}$&    ...          &5.7$\pm$0.3$^{1}$&3.7$\pm$0.2$^{13}$&4.4$\pm$0.2$^{1}$&     $<4.7$  $^{1}$&6.03$\pm$0.20$^{9}$\\
281 &     ...          &6.7$\pm$0.3$^{2}$&6.5$\pm$0.2$^{3}$&    ...          &    ...          &3.9$\pm$0.2$^{12}$&4.8$\pm$0.3$^{1}$&    ...            &6.14$\pm$0.20$^{7}$\\
354 &11.3$\pm$0.3$^{1}$&6.2$\pm$0.3$^{2}$&6.4$\pm$0.2$^{2}$&    ...          &5.4$\pm$0.3$^{1}$&4.3$\pm$0.2$^{5}$ &    ...          &    ...            &6.34$\pm$0.20$^{6}$\\
389 &     ...          &6.5$\pm$0.3$^{2}$&7.3$\pm$0.3$^{1}$&7.4$\pm$0.3$^{1}$&5.6$\pm$0.3$^{1}$&4.9$\pm$0.2$^{8}$ &    ...          &    ...            &7.90$\pm$0.20$^{14}$\\
363 &      $<9.4$$^{1}$&6.2$\pm$0.3$^{2}$&    ...          &    ...          &6.3$\pm$0.3$^{1}$&5.3$\pm$0.2$^{5}$ &    ...          &    ...            &8.00$\pm$0.20$^{13}$\\
392 &10.9$\pm$0.3$^{1}$&6.3$\pm$0.3$^{2}$&5.2$\pm$0.2$^{2}$&6.8$\pm$0.3$^{6}$&5.7$\pm$0.3$^{1}$&4.8$\pm$0.2$^{12}$&4.9$\pm$0.3$^{3}$&6.2$\pm$0.2$^{1}$  &7.93$\pm$0.20$^{48}$\\
434 &      $<9.3$$^{1}$&6.5$\pm$0.3$^{2}$&6.1$\pm$0.3$^{2}$&    ...          &6.0$\pm$0.3$^{1}$&5.1$\pm$0.2$^{6}$ &6.5$\pm$0.5$^{1}$&    ...            &7.90$\pm$0.20$^{24}$\\
469 & 9.8$\pm$0.3$^{1}$&6.7$\pm$0.3$^{2}$&6.6$\pm$0.2$^{1}$&    ...          &6.0$\pm$0.2$^{2}$&5.4$\pm$0.2$^{7}$ &5.8$\pm$0.5$^{2}$&6.5$\pm$0.3$^{1}$  &8.11$\pm$0.20$^{28}$\\
535 & 9.8$\pm$0.3$^{1}$&6.2$\pm$0.3$^{2}$&7.2$\pm$0.2$^{4}$&    ...          &5.8$\pm$0.3$^{1}$&4.6$\pm$0.2$^{1}$ &    ...          &    ...            &7.90$\pm$0.20$^{3}$\\
489 & 9.8$\pm$0.3$^{1}$&6.6$\pm$0.3$^{2}$&7.1$\pm$0.2$^{3}$&6.7$\pm$0.3$^{1}$&6.2$\pm$0.3$^{1}$&5.6$\pm$0.2$^{2}$ &7.4$\pm$0.3$^{1}$&    ...            &7.87$\pm$0.20$^{21}$\\
\hline
\end{tabular}
\label{t:tab5}
\end{table*}

\begin{table*}
\caption{Abundance relative to solar for stars in NGC~1904}
\begin{tabular}{lllllllllll}
\hline
\hline
star&   [He/H]  &    [Mg/H]  &   [Si/H]   &    [P/H]   &   [Ca/H]   &  [Ti/H]    &   [Cr/H]   &   [Mn/H]   & [Fe/H]    \\
\hline
209&     ...    &-1.1$\pm$0.2&-1.3$\pm$0.2&     ...    &-0.6$\pm$0.3&-1.3$\pm$0.2&-1.3$\pm$0.2&   $<-0.7$  &-1.48$\pm$0.20\\
281&     ...    &-0.8$\pm$0.2&-1.0$\pm$0.2&     ...    &     ...    &-1.1$\pm$0.2&-0.9$\pm$0.3&     ...    &-1.37$\pm$0.20\\
354&+0.3$\pm$0.3&-1.3$\pm$0.2&-1.1$\pm$0.2&     ...    &-0.9$\pm$0.3&-0.7$\pm$0.2&     ...    &     ...    &-1.17$\pm$0.20\\
389&     ...    &-1.0$\pm$0.2&-0.2$\pm$0.3&+1.9$\pm$0.3&-0.7$\pm$0.3&-0.1$\pm$0.2&     ...    &     ...    &+0.39$\pm$0.20\\
363&  $<-1.6$   &-1.3$\pm$0.2&    ...     &     ...    & 0.0$\pm$0.3&+0.3$\pm$0.2&     ...    &     ...    &+0.49$\pm$0.20\\
392&-0.1$\pm$0.3&-1.2$\pm$0.2&-2.3$\pm$0.2&+1.3$\pm$0.3&-0.6$\pm$0.3&-0.2$\pm$0.2&-0.8$\pm$0.3&+0.8$\pm$0.2&+0.42$\pm$0.20\\
434&  $<-1.7$   &-1.0$\pm$0.2&-1.4$\pm$0.2&     ...    &-0.3$\pm$0.3&+0.1$\pm$0.2&+0.8$\pm$0.5&     ...    &+0.39$\pm$0.20\\
469&-1.2$\pm$0.3&-0.8$\pm$0.2&-0.9$\pm$0.3&     ...    &-0.3$\pm$0.2&+0.4$\pm$0.2&+0.1$\pm$0.5&+1.1$\pm$0.3&+0.60$\pm$0.20\\
535&-1.2$\pm$0.3&-1.3$\pm$0.2&-0.3$\pm$0.2&     ...    &-0.5$\pm$0.3&-0.4$\pm$0.2&     ...    &     ...    &+0.39$\pm$0.20\\
489&-1.2$\pm$0.3&-0.9$\pm$0.2&-0.4$\pm$0.2&+1.2$\pm$0.3&-0.1$\pm$0.3&+0.6$\pm$0.2&+1.7$\pm$0.3&     ...    &+0.36$\pm$0.20\\
\hline
\end{tabular}
\label{t:tab6}
\end{table*}

\subsection{Abundance anomalies}

In this section we present the results of the abundance analysis.
Tables~\ref{t:tab5} and \ref{t:tab6} contain the abundance values
derived for each of our target stars, both on an absolute scale, and
relative to the solar abundances of Grevesse \& Sauval (1998). The
scatter of the measurements for the corresponding lines gives the
quantity $\sigma_{N}$, which is added quadratically to the error
contributions in $T_{\rm eff}$, $g$, $\xi$ and [A/H] described in the
previous chapter, to get the final error for each element in each
target star.

\subsection{Star 392}

\begin{figure}
\centering
\includegraphics[width=8.8cm]{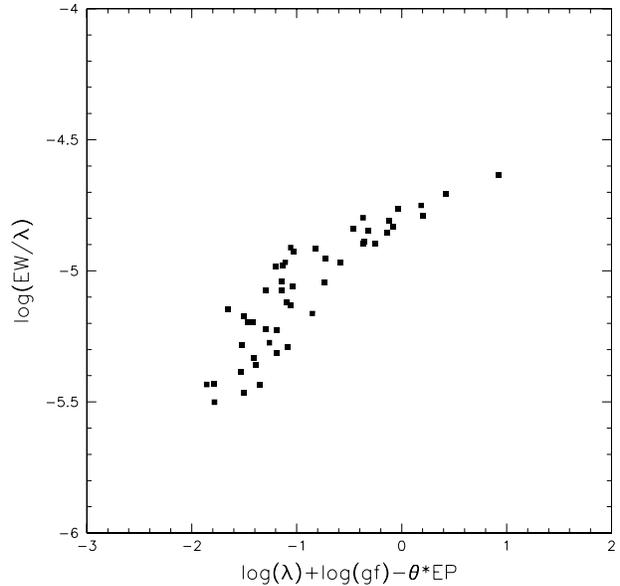}
\caption{Curve of growth for \ion{Fe}{ii} lines in the spectrum of star 392}
\label{f:fig5}
\end{figure}

\begin{figure}
\centering
\includegraphics[width=8.8cm]{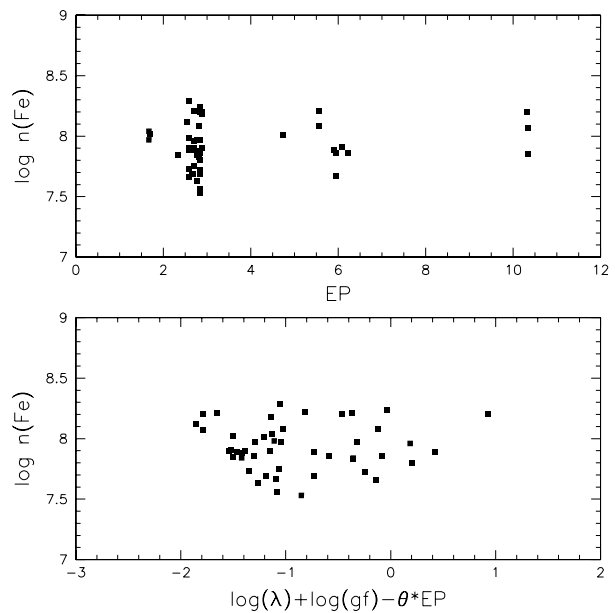}
\caption{Upper panel: run of the extracted abundances for individual \ion{Fe}{ii}
lines from the spectrum of star 392 with excitation potential (in
EV). Lower panel: the same, but against expected line strength}
\label{f:fig6}
\end{figure}

We will first consider the analysis of star 392, whose spectrum has
the highest $S/N$\ ($\sim 60$\ per resolution element at a wavelength
of 4240~\AA). For this star we were able to measure over 100 lines.

Figure~\ref{f:fig5} shows the curve of growth of \ion{Fe}{ii}.
Clearly, many unsaturated lines were accurately measured.  As shown in
Figure~\ref{f:fig6}, no trend was detected in the number of Fe atoms,
log~n(Fe), with the excitation potential, as expected for the
excitation equilibrium. This test is very critical, given the large
spread of excitation potentials (nearly 10 eV): even a modest error of
500~K in the effective temperature would result in a very significant
trend in this plot.

The large number of measured lines for \ion{Fe}{ii} allowed an
accurate estimate of the microturbulent velocity: best results were
obtained setting it to zero (i.e., leaving only thermal broadening as
an unsaturating factor). This result is not surprising in view of the
fact that this star is so hot that the atmosphere should be in
radiative equilibrium.

With our value of the effective temperature and surface gravity, we
got an excellent equilibrium of ionization for Fe: the mean abundance
given by neutral lines is log~n(Fe)=$7.86\pm 0.03$\ (31 lines, r.m.s.
of results for individual lines of 0.18~dex), while that obtained from
singly ionized lines is log~n(Fe)=$7.93\pm 0.03$\ (48 lines, r.m.s. of
results for individual line of 0.20~dex). The offset between the
abundances is only marginally larger than the sum of the internal
errors; again, a small change of the effective temperatures and
gravities (within the internal errors of both these quantities) would
bring the two abundances in agreement with each other.

This very good agreement supports our LTE analysis. Actually, such a
good agreement was not foreseen, since we expected some overionization
of Fe, leading to lower abundances from neutral lines in an LTE
analysis. This result should be compared with appropriate predictions
from statistical equilibrium calculations for these stars; however,
this is beyond the scopes of the present analysis.

\subsection{Other stars}

The abundance values for the various elements for which we have data
are plotted in Figures~\ref{f:fig8}, \ref{f:fig9},
\ref{f:fig10}, \ref{f:fig11}. In the upper panels, the abundances
for individual stars in NGC~1904 are plotted as a function  of
$T_{\rm eff}$\ as log offsets from the solar abundances. The
abundance values from Behr (2003a) for M13 are plotted  for
comparison in  the lower panels. The  horizontal lines represent the
expected value of [metal/H] from solar for NGC~1904 and M13 in  the
Carretta \& Gratton  (1997) scale: at -1.37 and -1.39, respectively.
For helium, the horizontal line represents the  solar ratio.  We
used the abundances derived   from lines of  the dominant ionization
stage of each element, to   minimize  the possibility of non-LTE
effects.

\subsubsection{Iron, titanium and chromium}

\begin{table}
\caption{Statistics of Fe Abundances}
\begin{tabular}{lllllllllll}
\hline
\hline
star &\multicolumn{3}{c}{\ion{Fe}{i}}&\multicolumn{3}{c}{\ion{Fe}{ii}}\\
     & n & $<>$ &$\sigma$ & n & $<>$ &$\sigma$ \\
\hline
 209 &29 & 6.06 & 0.19 &  9 & 6.03 & 0.25 \\
 281 &19 & 6.47 & 0.27 &  7 & 6.14 & 0.26 \\
 354 & 2 & 6.90 & 0.18 &  6 & 6.34 & 0.32 \\
 389 &11 & 7.66 & 0.27 & 14 & 7.90 & 0.31 \\
 363 & 8 & 8.24 & 0.47 & 13 & 8.00 & 0.38 \\
 392 &31 & 7.86 & 0.18 & 48 & 7.93 & 0.20 \\
 434 & 9 & 7.80 & 0.44 & 24 & 7.90 & 0.48 \\
 469 & 9 & 7.87 & 0.26 & 28 & 8.11 & 0.34 \\
 535 &   &      &      &  3 & 7.90 & 0.06 \\
 489 &   &      &      & 21 & 7.87 & 0.46 \\
\hline
\end{tabular}
\label{t:tab7}
\end{table}

\begin{figure}[!h]
\centering
\includegraphics[width=8.8cm]{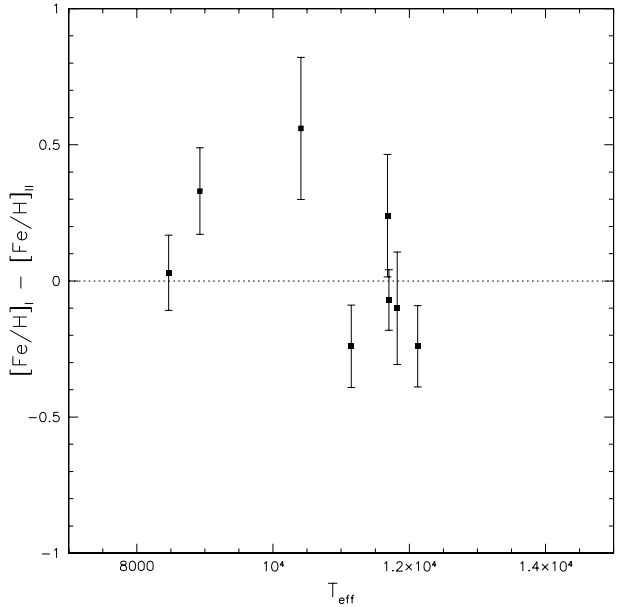}
\caption{Differences between abundances for \ion{Fe}{i} and
\ion{Fe}{ii} lines as a function of temperature}
\label{f:fig7}
\end{figure}

\begin{figure}
\includegraphics[width= 8.8 cm]{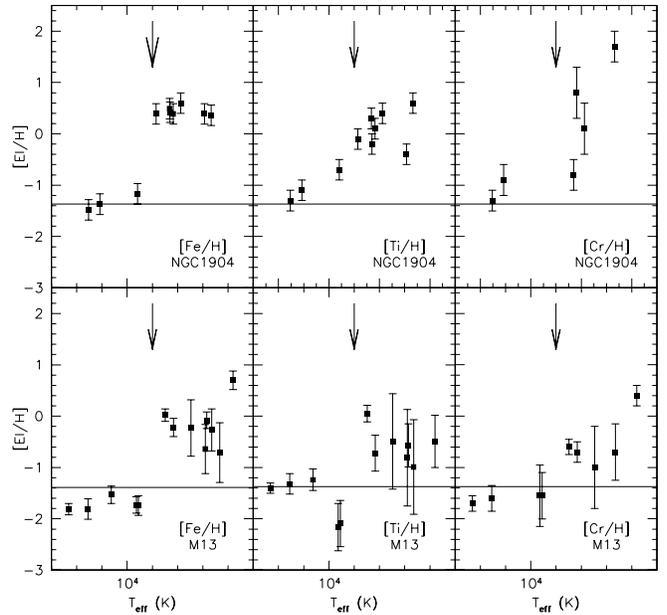}
\caption{Abundance values for iron, titanium and chromium as a
function of $T_{\rm eff}$ as log offsets from the solar abundances.
The abundance values from Behr et al. (2003a) for M13 are also
plotted for comparison in the lower panels. The horizontal solid
lines represent the expected value of [metal/H] from solar for
NGC~1904 and M13, in the Carretta \& Gratton (1997) scale. The
arrows at 11\,000 K indicate the approximate position of Grundahl's
photometric jump}
\label{f:fig8}
\end{figure}

Statistics of abundances for Fe are given in Table~\ref{t:tab7}.
Abundances are obtained from both neutral and singly ionized lines, in
a few cases from a total of more than 30 lines. Figure~\ref{f:fig7}
displays the run of the differences between the average abundances
obtained from neutral and singly ionized Fe lines as a function of
temperature.  Abundances from \ion{Fe}{i} lines agree well with those
from
\ion{Fe}{ii} lines. However,  given the possibility that small
non-LTE effects  are  present, in  the following  discussion we
considered the abundances  provided by singly ionized  Fe, which is
the dominant   species and   should   then   be  less  affected  by
model uncertainties.

As can be seen from Figure~\ref{f:fig8}, the iron abundances for the
three coolest stars are very close to the cluster metallicity from
analysis of red giant branch stars. We obtain [Fe/H]=$-1.48$, $-1.37$,
and $-1.17$, yielding an average value of [Fe/H]=$-1.34\pm 0.09$. This
value may be compared with those obtained from analysis of red giants:
[Fe/H]=-1.37 on the Carretta \& Gratton (1997) scale and the slightly
lower value of [Fe/H]=-1.59 given by Kraft \& Ivans (2003).

On the other hand we find remarkable enhancements of iron and other
metal species among the blue HB stars hotter than $\sim$ 11\,000
K. The coolest star showing signs of radiative levitation is star
389 at $T_{\rm eff} = 11\,148
\pm 200$~K, with a supra-solar value of [Fe/H]$=+0.39\pm 0.20$. Stars hotter
than $\sim$11\,000~K show     a  very  similar   iron    content:   the  average
value is [Fe/H]=$+0.43\pm 0.05$.   The very   small scatter (0.08~dex)  may  be
justified by  the internal errors in  the atmospheric parameters. Both
the threshold temperature  at which the  anomalous abundances appear,
as well  as  the amplitude  of  the  enhancement (about  1.8  dex) are
similar to those found by Behr et al. (2003a) for various clusters.

In a similar way, titanium is found a few tenths of a dex above the
cluster baseline in the cooler stars ([Ti/H]=-1.3, -1.1, and
-0.7). The average value of [Ti/Fe]=+0.31 agrees very well with the
overabundance of this element as generally found in cooler metal-poor
stars. Titanium abundances then rise by a factor quite similar to that
found for Fe in the hotter population.

A similar trend is also obtained for chromium, again in good agreement
with the results by Behr (2003a); however, Cr abundances are based on
only two \ion{Cr}{ii} lines.

\subsubsection{Helium}

\begin{figure}
\centering
\includegraphics[width=8.8cm]{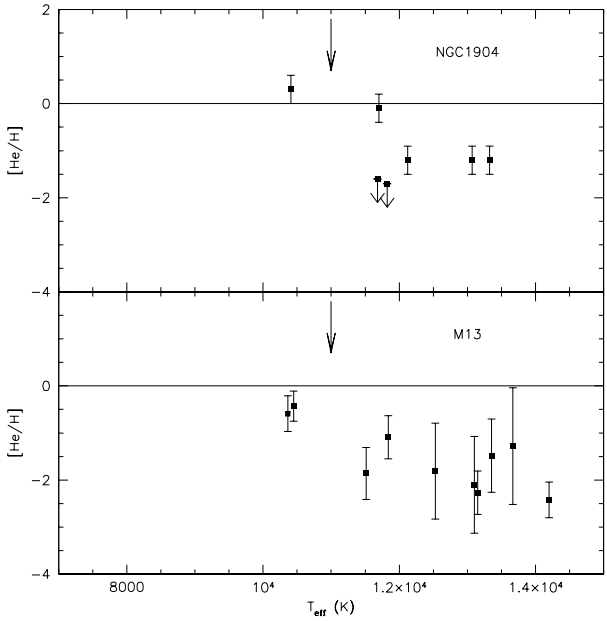}
\caption{Same as Figure \ref{f:fig8}, but for Helium}
\label{f:fig9}
\end{figure}

The He abundances appear below the expected\footnote{We neglect here
the difference between the solar and cosmological He abundance, much
smaller than the observational errors of our He abundance
determinations.} solar He/H ratio in the stars with $T_{\rm
eff}>11\,000$~K. However, at variance with the case of Fe, we found a
large and likely significant star-to-star scatter in the He
depletions, see Fig. ~\ref{f:fig9}. The case of star 392 is very
significant: the relatively high $S/N$\ allows clear detection of the
He lines. The He abundance we derive from the 4471~\AA\ line is
consistent with the solar value. On the other hand, no He line at all
was detected in other stars having a similar temperature.  Weak lines
were detected in the spectra of the warmer stars, suggesting a large
He depletion of at least an order of magnitude.

\subsubsection{Phosphorus and manganese}

\begin{figure}[!t]
\centering
\includegraphics[width=8.8cm]{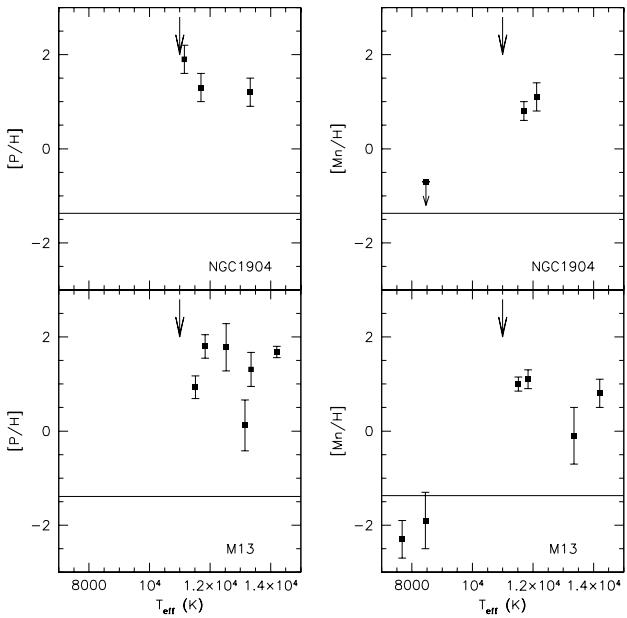}
\caption{Same as Figure \ref{f:fig8}, but for phosphorus and
manganese}
\label{f:fig10}
\end{figure}

Some metals, such as phosphorus and manganese (see
Fig.~\ref{f:fig10}), display significantly larger enhancements than
iron. We do not observe any \ion{P}{ii} line among the cooler stars,
but if we assume an appropriately-scaled solar composition for these
stars, then the supra-solar value of [P/H] $\simeq$ +1.5 that we find
for the hot stars implies an enhancement of $\sim$3 orders of
magnitude with respect to the cluster's adopted metallicity. It is
worth noting that several \ion{P}{ii} lines were found in the field HB
stars Feige 86 and 3 Cen A by Sargent
\& Searle (1967) and Bidelman (1960),  respectively, suggesting that
the same mechanism may  be at work in  both cluster stars  and field
stars.  On the other hand, Behr (2003a) found very similar P
enhancements  for hot HB stars in    M13. Similar very large
overabundances   are also observed for manganese.

\subsubsection{Magnesium, silicon and calcium}

\begin{figure}
\includegraphics[width=8.8cm]{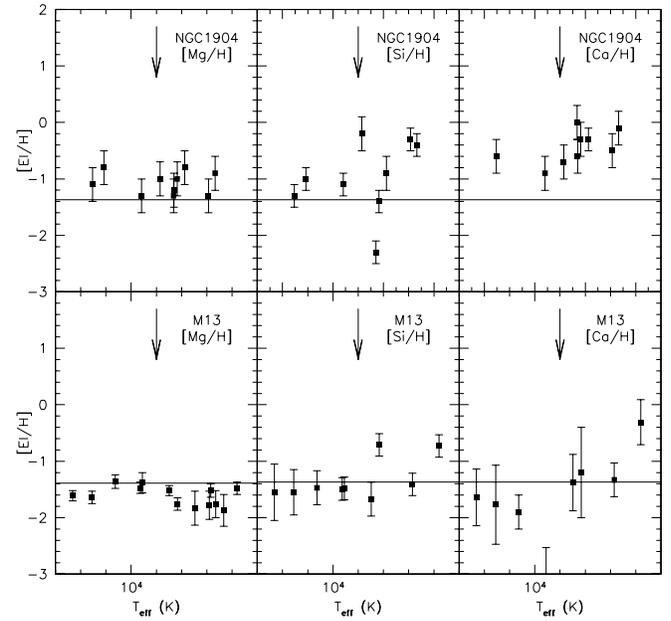}
\caption{Same as Figure \ref{f:fig8}, but for magnesium, silicon and calcium}
\label{f:fig11}
\end{figure}

Figure~\ref{f:fig11} displays the run of the abundances of magnesium,
silicon and calcium with temperature. The abundance of magnesium was
obtained from the strong multiplet of \ion{Mg}{ii} at 4481~\AA, that
is partially resolved in our spectra; Ca abundance was derived from
the resonance K-line.

The abundances for magnesium, silicon and calcium, both below and
above the critical temperature of 11\,000~K, are consistent with very
little or no enhancement, but with a large, likely real scatter for
Si. This result agrees with that found by Behr (2003a) and with
theoretical expectations.

\section{Discussion and conclusions}

The abundance anomalies in NGC~1904 are likely to be due to the same
diffusion processes that were invoked by Behr et al.\ (1999):
radiative levitation of metals and gravitational settling of helium,
in the stable non-convective atmospheres of the hotter, higher gravity
stars, as hypothesized by Michaud et al. (1983). In addition, the jump
in the $u$,$(u-y)$ diagram by Grundahl et al.\ (1999) for NGC~1904 is
reported at $\log T_{\rm eff}=4.06\pm 0.01$, in very good agreement
with the onset of the abundance anomalies observed here.

An underabundance of helium on the BHB was observed in several
previous instances (Baschek 1975; Heber 1987; Glaspey et al.  1989;
among others) and in fact appears to be typical of stars of this
type. Michaud et al. (1983), building on the original suggestion by
Greenstein et al. (1967), explained these underabundances as a result
of the gravitational settling of helium, which can take place if the
outer atmosphere of the star is sufficiently stable. Our current
results confirm a distinct trend in [He/H] with $T_{\rm eff}$ and
$\log g$\ along the HB.  Helium depletion is accompanied by
photospheric enhancement of metals, since the same stable atmosphere
that permits gravitational settling also allows the levitation of
elements with large radiative cross sections. Elements which have
sufficiently large cross-sections to the outgoing radiation field will
experience radiative accelerations greater than gravity and will
diffuse upwards, enriching the photosphere. The gross behaviour of the
diffusion models can be understood by considering the migration of
chemical species in a stellar atmosphere (Gonzalez et al. 1995;
Michaud et al.  1983 and Seaton 1997).  The relative amount of
levitation experienced by each element can be estimated by summing the
predicted equivalent widths for each line over the visible and near UV
spectral range.  The magnitudes of these net equivalent widths suggest
that helium will experience only a weak levitation force, insufficient
to oppose gravity, magnesium will be levitated by an intermediate
amount, perhaps balanced by gravity, while iron will be strongly
levitated. This is admittedly a crude approach to what we see in
NGC~1904, and Behr et al.\ measurements in various clusters.

The identification of the detected spectral lines was successful in
most cases. A few weak lines were not identified, which could be due
to the low S/N of the spectra. Since the abundances of metals we
derived are very peculiar, it is possible that other metal species
(usually not observed in late B-type stars) are also present in the
atmospheres of these hot BHB stars, but based on our data we cannot
draw any conclusions about this. Very weak lines were used only for
the star having the best spectrum. Our results thus do not change if
we remove the abundance determinations from those weak lines.

In general, our abundance measurements in NGC~1904 give very similar
results to those obtained previously by Behr et al.\ (1999) in M13,
which, on the other hand, has a very similar metallicity to that of
NGC~1904. The very small scatter in Fe abundances among the warmer
stars is particularly intriguing. While it could be the result of
small statistics, it might indicate that the stars warmer than
11\,000~K observed in NGC~1904 all rotate at very similar rates.  In
fact, Michaud (2002, private comunication) estimates that small
differences of about 2-3 km/s in rotational velocity could generate
sizeable discrepancies in the radiative levitation and gravitational
settling processes.  Unfortunately, the accuracy of our rotational
measurements in NGC~1904, with errors of the order of 4-5 km/s, do not
allow us to explore this possibility.

\bigskip
\begin{acknowledgements}
This research has made use of the NIST database, operated by the
National Institute of Standards and Technology.  We thank F. Grundahl
and Y. Momany for gently providing their photometries and B. B. Behr
for useful discussion and helpful suggestions and for sharing his
results on abundance measurements prior to pubblication. We also thank
the anonymous referee for valuable comments that have improved this
paper. DF acknowledges support provided by PRIN2001 (Italian Ministero
dell'Istruzione e della Ricerca).  ARB recognizes the support of the
{\it INAF}.  GP and RG recognize partial support from the {\it MIUR}
(COFIN 2001028897) and from the {\it ASI}.
\end{acknowledgements}


\begin{thebibliography}{}

\bibitem[]{}Bard, A., \& Kock, M. 1994, A\&A, 282, 1014
\bibitem[]{}Baschek, B. 1975, in {\it Problems in stellar atmospheres and envelopes}, Springer, New York, p.101
\bibitem[]{}Behr, B. B. 2003a, ApJS, 149, 67
\bibitem[]{}Behr, B. B. 2003b, ApJS, 149, 101
\bibitem[]{}Behr, B. B., Cohen, J. G., McCarthy, J. K., \& Djorgovski, S. G. 1999, ApJL, 517, L31
\bibitem[]{}Behr, B. B., Djorgovski, S. G., Cohen, J. G., et al. 2000, ApJ, 528, 849
\bibitem[]{}Behr, B. B., Cohen, J. G., \& McCarthy, J. K. 2000b, ApJL, 531, L37
\bibitem[]{}Bidelman, W. P. 1960, PASP, 72, 24
\bibitem[]{}Biemont, E., Baudoux, M., Kurucz, R. L., Ansbacher W. \&
   Pinnington, E. H. 1991, A\&A, 249, 539
\bibitem[]{}Bragaglia, A., Carretta, E., Gratton, R. G., et al. 2001, AJ, 121,
   327
\bibitem[]{}Brown, T. M., Bowers, C. W., Kimble, R. A., \& Ferguson, H. C. 2000,
   ApJL, 529, L89
\bibitem[]{}Brown, T. M., Sweigart, A. V., Lanz, T., Landsman, W. B., \& Hubeny, I. 2001, ApJ, 562, 368
\bibitem[]{}Carretta, E., \& Gratton, R. G. 1997, A\&AS, 121, 95
\bibitem[]{}Cassisi, S., Castellani, V., degl'Innocenti, S., Salaris, M., \&
   Weiss, A. 1999, A\&AS, 134, 103
\bibitem[]{}Castelli, F., Gratton, R. G., Kurucz, R. L. 1997, A\&A, 318, 841
\bibitem[]{}Castelli, F., Gratton, R. G., Kurucz, R. L. 1997, A\&A, 324, 432
\bibitem[]{}Cayrel, R. 1988, in {\it The Impact of Very High S/N Spectroscopy on
   Stellar Physics}, IAU Symp. 132, eds. G. Cayrel de Strobel \& M. Spite,
   Kluwer, Dordrecht, p. 345
\bibitem[]{}Crocker, D. A., Rood, R. T., O'Connell, R. W. 1988, ApJ, 332, 236
\bibitem[]{}D'Antona, F., Caloi, V., Montalban, J., Ventura, P., \& Gratton, R. 2002, A\&A, 395, 69
\bibitem[]{}Dixon, W. V., Davidsen, A. F., Dorman, B., Ferguson, H. C. 1996, AJ 111, 1936
\bibitem[]{}Glaspey, J. W., Michaud, G., Moffat, A. F. \& Demers, S. 1989, ApJ,
   339, 926
\bibitem[]{}Gonzalez, J.-F., Artru, M.-C., \& Michaud, G. 1995, A\&A, 302, 788
\bibitem[]{}Gratton, R. G., Carretta, E., Claudi, R., Lucatello, S., \&
   Barbieri, M. 2003, A\&A, 404, 187
\bibitem[]{}Greenstein, G. S., Truran, J. W. \& Cameron, A. G. W. 1967, Nature,
    213, 871
\bibitem[]{}Grevesse, N., \& Sauval, A. J. 1998, Space Science Reviews, 85, 161
\bibitem[]{}Grundahl, F., Vandenberg, D. A., \& Andersen, M.I. 1998, ApJ, 500,
    L179
\bibitem[]{}Grundahl, F., Catelan, M., Landsman, W. B., Stetson, P. B., \&
    Andersen, M. I. 1999, ApJ, 524, 242
\bibitem[]{}Hambly, N. C., Rolleston, W. R. J., Keenan, F. P., Dufton, P. L.,
    \& Saffer, R. A. 1997, ApJS, 111, 419
\bibitem[]{}Harris, W. E. 1996, AJ, 112, 1487
\bibitem[]{}Heber, U. 1987, Mitt. Astron. Ges., 70, 79
\bibitem[]{}Heise, C., \& Kock, M. 1990, A\&A, 230, 244
\bibitem[]{}Hui-Bon-Hoa, A., LeBlanc, F., \& Hauschildt, P. H. 2000, ApJ, 535, L43
\bibitem[]{}Kaufer, A., D'Odorico, S., \& Kaper, L. 2003, {\it UV-visual Echelle
   Spectrograph User Manuali}, Doc. No. VLT-MAN-ESO-13200-1825, Issue 1.7,
   ed. S. Hubrig \& C. Ledoux
\bibitem[]{}Kraft, R. P. 1994, PASP, 106, 553
\bibitem[]{}Kraft, R. P., \& Ivans, I. I. 2003, PASP, 115, 143
\bibitem[]{}Kurucz, R. 1994, CD-ROM 19, Smithsonian, Cambridge
\bibitem[]{}Kurucz, R., \& Bell, B. 1995, CD-ROM 23, Smithsonian, Cambridge
\bibitem[]{}Mengel, J. G. \& Gross, P. G. 1976, Ap\&SS, 41, 407
\bibitem[]{}Michaud, G., Vauclair, G., \& Vauclair, S. 1983, ApJ, 267, 256
\bibitem[]{}Moehler, S., Heber, U., \& DeBoer, K.S. 1995, A\&A, 294, 65
\bibitem[]{}Moehler, S., Heber, U., Rupprecht, G. 1997, A\&A, 319, 109
\bibitem[]{}Moehler, S., Sweigart, A. V., Landsman, W. B., Heber, U., \& Catelan, M. 1999, A\&A, 346, 1
\bibitem[]{}Moehler, S., Sweigart, A. V., Landsman, W. B., \& Heber, U. 2000,
   A\&A, 360, 120
\bibitem[]{}Moehler, S., Landsman, W. B., Sweigart, A. V., \& Grundahl, F. 2003,
   A\&A, 405, 135
\bibitem[]{}Momany, Y., Bedin, L. R., Cassisi, S., Piotto, G., Ortolani, S., Recio-Blanco, A., De Angeli, F., \& Castelli, F. 2004, A\&A, 420, 605 %
\bibitem[]{}Moore, C. E., Minnaert, M. G. J., \& Houtgast, J. 1966, NBS Monog.,
   Washington
\bibitem[]{}Peterson, R. C. 1983, ApJ, 275, 737
\bibitem[]{}Peterson, R. C. 1985a, ApJ, 289, 320
\bibitem[]{}Peterson, R. C. 1985b, ApJ, 294, 35
\bibitem[]{}Peterson, R. C., Tarbell, T. D., \& Carney, B. W. 1983, ApJ, 265, 972
\bibitem[]{}Recio-Blanco, A., Piotto, G., Aparicio, A., \& Renzini, A. 2002,
    ApJ, 572, L71
\bibitem[]{}Richer, J., Michaud, G., \& Turcotte, S. 2000, ApJ, 529, 338
\bibitem[]{}Sandage, A. R., \& Wildey, R. 1967, ApJ, 150, 469
\bibitem[]{}Sargent, W. L. W. \& Searle, L. 1967, ApJL, 150, L33
\bibitem[]{}Schlegel, D. J., Finkbeiner, D. P., \& Davis, M. 1998, ApJ, 500, 525
\bibitem[]{}Seaton, M. J. 1997, MNRAS, 289, 700
\bibitem[]{}Sills, A., \& Pinsonneault, M. H. 2000, ApJ, 540, 489
\bibitem[]{}Simmons, G. J., \& Blackwell, D. E. 1982, A\&A, 112, 209
\bibitem[]{}Sweigart, A. V. 2002, in {\it Highlights of Astronomy}, Vol 12, ed.
    H. Rickman, ASP, S. Francisco, p. 292
\bibitem[]{}Sweigart, A. V., \& Catelan, M. 1998, ApJ, 501, 63
\bibitem[]{}Turcotte, S., Richer, J., \& Michaud, G. 1998, ApJ, 504, 559
\bibitem[]{}van den Bergh, S. 1967, AJ, 72, 70

\end{thebibliography}
\end{document}